\documentclass[twocolumn,prl]{revtex4}
\usepackage{graphicx}
\usepackage{amsmath}

\begin{document}
\title{Non-exponential tunneling and control of microwave absorption
lineshapes via Fano resonance for electrons on Helium}
\author{Solomon Duki}
\author{Harsh Mathur}
\affiliation{Department of Physics, Case Western Reserve University,10900 Euclid Avenue, Cleveland OH 44106-7079}

\begin{abstract}

We consider the application of a small in-plane magnetic field 
to electrons on a helium surface in a perpendicular magnetic
field. Certain states that were bound to the helium surface then
dissolve into the continuum turning into long-lived resonances.
As a result microwave absorption lines acquire an 
asymmetric Fano lineshape  that is tunable by varying the
microwave polarisation or the in-plane magnetic field. 
Electrons trapped in a formerly
bound state will tunnel off the surface of helium; we show
that under suitable circumstances this ``radioactive decay'' can 
show damped oscillations rather than a simple exponential
decay. The mechanism for oscillatory exponential decay
is not specific to electrons on Helium and this effect 
may also be relevant elsewhere in physics.

\end{abstract}

\maketitle

Electrons deposited on the surface of a pool of liquid helium form a 
high mobility two-dimensional electron gas that has been used
to study fundamental problems of condensed matter physics including
electron localisation and Wigner crystallisation \cite{cole}. Recently there has been
interest in using this system to build a quantum computer due to its high
degree of quantum coherence \cite{dykman}. In this design it is envisaged that
quantum computing protocols will be implemented by driving transitions
between electronic subband states using microwaves. 

The purpose of this Letter is to show that in the presence of a
strong magnetic field, tilted with respect to the helium surface, the 
shape of some microwave absorption lines is determined by Fano 
resonance effects \cite{fano}. This allows the lineshape to be precisely controlled 
by varying the magnetic field and microwave polarisation. Apart from any
practical relevance, using quantum interference to control lineshapes is
also of fundamental interest. For example, electromagnetically induced
transparency is a quantum interference effect that alters lineshapes \cite{eim};
its implementation in Bose condensates has led to fascinating phenomena,
such as an extreme slowing down of light \cite{hau}.

The key physics that underlies
the effects discussed here is that in a perpendicular magnetic field it is
possible for certain electronic states that are bound to the surface of helium 
to be nevertheless degenerate with unbound continuum states. When the
the magnetic field is tilted, these bound states dissolve into the continuum
leaving behind long-lived resonances. Thus electrons that occupy the formerly
bound states will eventually escape the surface of helium via quantum tunneling. 
Remarkably we find that under suitable conditions this ``radioactive decay'' can
be non-exponential in its time dependence: the probability that the electron is 
still bound can undergo (damped) oscillations, again due to quantum 
interference. Such non-exponential tunneling decay may be relevant to
other areas of physics since the mechanism is not unique to electrons on
helium. Indeed there is interest in non-exponential quantum tunneling as far 
afield as cosmology where there has been a recent analysis of 
departures from exponential behaviour in the long-time limit 
\cite{lmk}.
The effect analysed here applies even in the short-time limit.
Another more formal result derived in this Letter is that the dissolved 
bound state leaves its imprint on the scattering phase shift of the 
continuum states; namely, the phase shift jumps by $\pi$ as
the energy is swept through the (renormalised) 
energy of the former bound state.

Before presenting our results it is useful to review Fano's analysis \cite{fano}
of a single bound state $| b \rangle$ that is coupled by a perturbation
$V$ to a continuum of states $|E \rangle$. The bound and the 
continuum states are taken to be eigenstates of an unperturbed
Hamiltonian, $H_0$, with eigenvalues $ \varepsilon_b$ and $E$ respectively.
It is assumed that the bound state is normalised,
$ \langle b | b \rangle = 1$, and orthogonal to the 
continuum states, $\langle b | E \rangle = 0$, which in turn are 
mutually orthonormal, $ \langle E | E' \rangle = \delta ( E - E' )$. 
The perturbation couples the bound state to 
the continuum states via the matrix element $ \langle E | V | b \rangle = V(E)$;
other matrix elements of the perturbation are presumed to vanish
[$\langle E | V | E' \rangle = \langle b | V | b \rangle = 0$].
As a result of the perturbation the bound state dissolves and the 
full Hamiltonian $H_0 + V$ only has continuum eigenstates, denoted
$ | \psi_E \rangle$, with continuous eigenvalue $E$. Fano's original
evaluation of the perturbed eigenstates $| \psi_E \rangle$ made use of arcane
delta function identities but in fact the same results may be
derived by observing that the state $|\psi_E \rangle$ must obey the
familiar Lipmann-Schwinger equation,
\begin{equation}
| \psi_E \rangle = 
| E \rangle + (E - H_0)^{-1} V | \psi_E \rangle,
\label{eq:lipmann}
\end{equation}
which can be solved iteratively to obtain the overlap
\begin{equation}
\langle b | \psi_E \rangle = \frac{1}{\pi V(E)} \sin [ \Delta(E) ]
\label{eq:overlap}
\end{equation}
and a more complicated expression for $\langle E' | \psi_E  \rangle$
\footnote{Some details: The first-order solution to eq (\ref{eq:lipmann}) is
obtained by replacing the perturbed eigenstate $| \psi_E \rangle$ 
on the right hand side with the
unperturbed state $|E\rangle$; the second-order solution,
by replacing $|\psi_E \rangle$ on the right-hand side with the first-order
solution; and so on. The resulting perturbative series is geometric. 
Its summation is facilitated by defining
the self energy $ \Sigma (E) = \langle b | V (E - H_0)^{-1} V |b \rangle$
by analogy to Dyson's equation in field theory. 
By this procedure we obtain eq (\ref{eq:overlap}) and 
$\langle E' | \psi_E \rangle =  \delta(E - E') \cos \Delta + (1/\pi)
[V(E')/V(E)] \sin \Delta /(E - E')$ in agreement with Fano's results.}.
Together these overlaps fully determine the perturbed eigenstate $| \psi_E \rangle$.
Here the phase angle $\Delta(E)$ is defined via
\begin{equation}
\tan \Delta(E) = \frac{ \pi | V(E) |^2 }{E - \varepsilon_b - F(E)}
\label{eq:delta}
\end{equation}
and $F(E)$ is the real part of the bound state self-energy
\begin{equation}
F(E) = P \int d E' | V(E') |^2 \frac{1}{E - E'}.
\label{eq:f}
\end{equation}
To the extent that we may assume $F(E)$ and $V(E)$ are slowly
varying, $ | \langle b | \psi_E \rangle |^2 $ is a sharply
peaked Lorentzian centred at $\epsilon_b$, the renormalised value
of the bound state energy (defined as the energy at which 
$E - \varepsilon_b - F(E)$ vanishes). The phase $\Delta(E)$ 
varies monotonically from $\pi$ to zero as $E$ varies from well 
below the renormalised energy,
$\epsilon_b$, to well above; it equals $\pi/2$ right at 
$\epsilon_b$.

Now suppose a time-dependent perturbation $T$ drives a transition
between some additional state $|i\rangle$ and the bound state 
$|b\rangle$ and the continuum states $|E\rangle$. A simple 
calculation using Fermi's golden rule shows that 
the transition rate to an exact eigenstate of energy
$E$ (normalised to the rate into the corresponding unperturbed
state) is given by
\begin{equation}
\frac{| \langle \psi_E | T | i \rangle |^2}{| \langle E | T | i \rangle |^2} 
= 
\frac{ | q + \epsilon |^2 }{1 + \epsilon^2}.
\label{eq:fano}
\end{equation}
Here the asymmetry parameter 
\begin{equation}
q = \frac{1}{ \pi V(E)^{\ast}} 
\frac{ \langle b | T | i \rangle }{ \langle E | T | i \rangle}
\label{eq:q}
\end{equation}
compares the relative couplings to the bound and unperturbed
continuum states and $\epsilon = \cot \Delta(E)$ is a suitably normalised 
measure of the distance in energy from location of the
Fano resonance, $\epsilon_b$. Eq (\ref{eq:q}) is the celebrated
lineshape derived by Fano. It shows that upon adding the perturbation
$V$, although the bound state dissolves into the continuum, it leaves
behind a trace in the form of a sharp absorption feature in transitions to 
the perturbed continuum. The Fano lineshape is asymmetric for generic $q$;
it reduces to the familiar symmetric Lorentzian or Breit-Wigner form only
in the limit $q \rightarrow \infty$. 

\begin{figure}
\scalebox{0.4}{\includegraphics{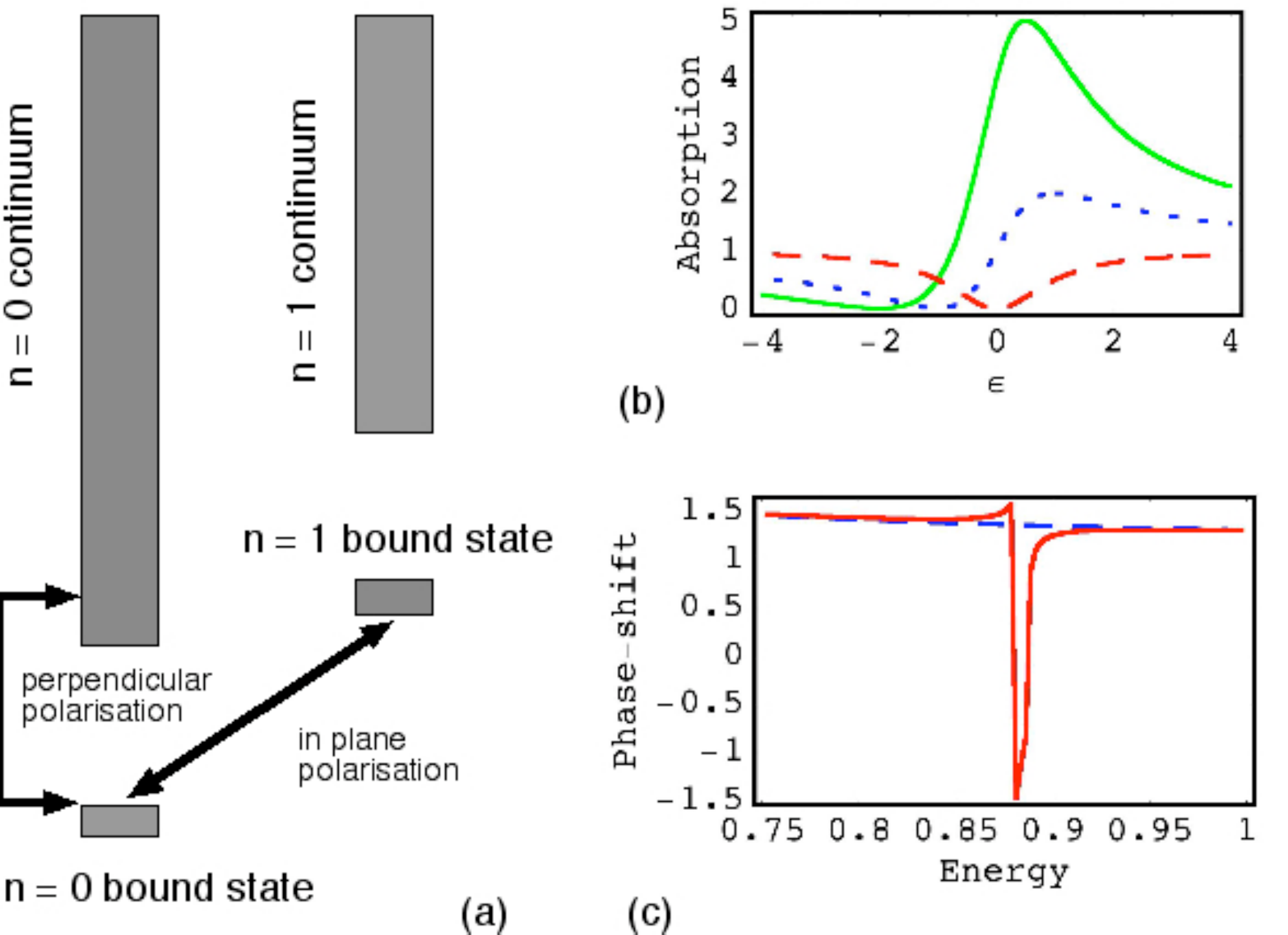}}
\caption{(a) Schematic energy level diagram for electrons on helium in a
perpendicular magnetic field. For simplicity only the lowest bound state
and the continuum states are shown for the $n=0$ and $n=1$ Landau 
levels. Also shown are the polarisations of microwave radiation needed to
drive transitions from the $n=0$ bound state to the $n=0$ 
continuum and the $n=1$ bound state. (b) The Fano lineshape for asymmetry
parameter $q = 0$ (dashed red curve), $q = 1$ (dotted blue curve) and $q = 2$
(solid green curve). (c) The phase shift for perturbed continuum states (solid red curve) calculated
by numerical solution of the full Schr\"{o}dinger equation shows the jump of $\pi$
relative to the unperturbed phase shift (dotted blue curve) as
predicted by the approximate truncation to Fano's model.}
\label{fig:plots}
\end{figure}

We turn now to the mapping between electrons on helium and
Fano's model. We take the helium surface to lie in the $y-z$ plane.
The electrons are bound to the surface by a potential
$V(x) = - Qe/4 \pi \epsilon_0 x + e F x$
for $ x > 0$. We assume that electrons cannot penetrate the helium
and apply a hardwall boundary condition at $x = 0$. 
Here $e$ is the magnitude of the electron charge and $Q = 7 \times 10^{-3} e$ is the
magnitude of the image charge that forms in the dielectric helium.
$F$ is an additional holding field that may be applied. In addition the
magnetic field is given by $B_x = B_{\perp}$, $B_y = 0$ and $B_z = B_{\parallel}$.
We adopt the gauge $A_x = - B_{\parallel} y$, $A_y = 0$ and $A_z = B_{\perp} y$.
It will be convenient to work in units where $ e B_{\perp} = \hbar = m = 1$ where
$m$ is the electron mass. The case that the inplane field vanishes corresponds 
to the unperturbed problem in Fano's model. In this case the
Schr\"{o}dinger equation is separable and has solutions of the form
$\exp(i p z) \varphi_{n}( y - p ) \xi(x)$ where $\varphi_n$ denotes the 
n$^{{\rm th}}$ eigenfunction of a one dimensional harmonic oscillator
of unit mass and frequency. The $x$ motion is governed by the 
one-dimensional equation
\begin{equation}
- \frac{1}{2} \frac{ \partial^2 }{ \partial x^2} \xi + V(x) \xi = 
\left[ E - \left(n + \frac{1}{2} \right) \right] \xi.
\label{eq:ximotion}
\end{equation}
Depending on whether the energy for $x$ motion, $ k^2/2 = E - (n + 1/2)$, is
negative or positive the wavefunction will be a bound subband wavefunction
$\xi_{\nu}(x)$ or a scattering state $\xi_{{\rm sc}}( x, k )$. In the absence of a
holding field $F$ the bound wavefunctions have the same form as the
s-wave states of hydrogen and a Bohr spectrum $ \varepsilon_{\nu} = - R/\nu^2$ 
where $\nu = 1, 2, 3,\ldots$ is the subband index of the bound state and $R \approx 
7.6$K.
The scattering states have the asymptotic form
\begin{equation}
\xi_{{\rm sc}} ( x, k ) \rightarrow  \frac{1}{\sqrt{2 \pi k}} 
\left( \exp[ - i k x - i \delta(k) ] + \exp[ i k x + \delta(k) ] \right)
\label{eq:asymptotics}
\end{equation}
as $ x \rightarrow \infty$.
By virtue of the hard wall boundary condition the reflected amplitude has
unit magnitude and the scattered state is fully specified by the single 
phase shift $\delta(k)$ between the incoming and outgoing states. 
In summary the eigenstates may be written
in the form $| p, n, \nu \rangle$ for the bound subband 
states and $| p, n, k \hspace{1mm}
({\rm sc}) \rangle$ for the unbound scattering states. The energy of the state is
given by $ E = (n + 1/2) + \varepsilon_{\nu} $ for the bound states and 
$ E = (n + 1/2) + k^2/2$ for the continuum. 
Note that the energy is independent of $p$ reflecting
the massive Landau degeneracy of these subband states. 

Fig 1 shows these energy levels schematically. From this figure we see that
for sufficiently large perpendicular field (of order 6 T) 
the lowest bound state in the $n=1$
Landau level can be degenerate with the unbound continuum states of the
$n=0$ Landau level. If we now turn on a small inplane magnetic field the 
most important effect of this perturbation on the bound state is to couple
it into the $n=0$ continuum \footnote{More
precisely, the bound state with a given $p$ gets coupled to the continuum state
with the same $p$ value.}. 
In the same way that an atom driven near 
resonance may be truncated into a two-level system, here we may truncate
the Hilbert space to just the $n=1$ bound state and the $n=0$ continuum.
This truncation maps the problem to Fano's model: the lowest
bound $n=1$ subband state
is identified with $|b\rangle$ and the $n=0$ continuum with $|E\rangle$.
In our units and gauge the perturbation is $ - i (B_{\parallel}/B_{\perp}) 
y \partial/\partial x$,
and the matrix element $V(E)$ in Fano's model is given by
\begin{equation}
V\left( \frac{k^2}{2} \right) = 
- \frac{i}{\sqrt{2}} \frac{B_{\parallel}}{B_{\perp}} \int_0^{\infty} d x \hspace{1mm}
\xi_b (x) \frac{\partial}{\partial x} \xi_{{\rm sc}} (x, k)
\label{eq:matrixmap}
\end{equation}

Having mapped the problem to Fano's model we now consider 
microwave absorption. For the unperturbed problem with a perpendicular
magnetic field it is easy to show that radiation that is polarised in the plane
couples a state $| p, n, \nu \rangle$ to states with the same $p$ and subband
state and Landau index $n \pm 1$. Radiation polarised perpendicular to the
plane leaves $p$ and the Landau index unchanged but can cause transitions
between the subbands or between a bound subband and a continuum state.
More explicitly the transition matrix element for radiation polarised in the
plane is
\begin{equation}
\langle p' n' \nu' | T_{\parallel} | p n \nu \rangle = i \delta( p - p' ) 
\frac{1}{\sqrt{2}} \frac{\Omega_E}{\omega} \delta_{\nu' \nu} 
(\delta_{n', n+1} + \delta_{n', n-1}).
\label{eq:paralleltransition}
\end{equation}
Here $\Omega_E = e {\cal E}/(\sqrt{\hbar e B_{\perp}})$ 
where ${\cal E}$ is the magnitude of the 
oscillating electric field and $\omega$ is its frequency. 
Similarly the transition matrix element for
radiation polarised perpendicular to the plane is
\begin{equation}
\langle p' n' \nu' | T_{\perp} | p n \nu \rangle = i \delta( p - p' ) 
\frac{\Omega_E}{\omega} \delta_{n' n} 
\int_0^{\infty} d x \hspace{1mm} \xi_{\nu'}^{\ast} (x) \frac{\partial}{\partial x} 
\xi_{\nu}(x).
\label{eq:perpendiculartransition}
\end{equation}

Now if we take the lowest subband state in the $n=0$ Landau level
as our state $|i \rangle$ evidently it is coupled to the state $|b\rangle$
(lowest subband state in the $n=1$ Landau level) by microwaves that
are polarised in the plane; it is coupled to the continuum $|E\rangle$
(unbound states in the $n=0$ Landau level) by microwaves polarised
perpendicular to the plane. Microwaves with intermediate polarisation
will couple to both bound state and continuum with a relative strength
that is tunable by varying the polarisation.
It follows that in the presence of a perturbing
inplane magnetic field the microwave absorption will show a Fano resonance
with an asymmetry parameter $q$ that can be tuned by varying polarisation 
from zero (pure perpendicular polarisation) to $\infty$ (pure in-plane polarisation).

Another consequence of our mapping to Fano's model is that when the
inplane perturbation is turned on the bound state $|b\rangle$ will dissolve
into the continuum $|E\rangle$ yielding the perturbed eigenstates $| \psi_E \rangle$.
Using Fano's expression for the overlap of $|\psi_E \rangle$ with $|E\rangle$ 
we find by straightforward asymptotic integration 
that the wavefunctions of the corresponding eigenstates have the
$x \rightarrow \infty$ asymptotic behaviour
\begin{eqnarray}
& \frac{1}{\sqrt{2 \pi k}} \exp(i p z) \phi_0 (y - p) 
( \exp[ - i k x - i \delta(k) - i \Delta ] &
\nonumber \\
& +  \exp[ i k x + i \delta(k) + i \Delta ] ). & 
\label{eq:phaseshift}
\end{eqnarray}
Comparing to the asymptotic $x$ dependence of the unperturbed 
states $|E\rangle$, eq (\ref{eq:asymptotics}),
we arrive at the elegant conclusion that although the bound
state dissolves into the continuum it leaves behind an imprint in
the perturbed continuum states in the form of an extra phase 
shift $\Delta$. As noted above the phase $\Delta$ jumps abruptly
by $\pi$ as we sweep through the renormalised energy of the
formerly bound state. 

We have derived this phase shift within the mapping 
to Fano's model. Since the prediction does not depend
on the specific form of the binding potential $V(x)$ we 
can check our prediction, and the veracity of the mapping
to Fano's model, by numerically solving the Schr\"{o}dinger
equation for the case that the confining potential is a
rectangular well, a circumstance that may be efficiently solved by
the numerical methods of N\"{o}ckel and Stone \cite{jens}. 
Fig 1 shows that the expected phase jump indeed occurs
without making the truncation to Fano's model.

Finally we turn to non-exponential tunneling decay.
As a prelude consider the basic Fano model of a single
bound state coupled by a perturbation to a continuum.
If the system starts in the bound state initially, the 
amplitude to remain in the bound state at time $t$ is
\begin{eqnarray}
b(t) & =  & \int d E \hspace{1mm} | \langle b | \psi_E \rangle |^2 
\exp[ - i E t ] 
\nonumber \\
& \approx & \exp( - i \epsilon_b t ) \exp \left( - \frac{t}{2 \tau_b} \right).
\label{eq:decay}
\end{eqnarray}
Here $\epsilon_b$ is the renormalised bound state energy defined
below eq (\ref{eq:f}) and $\tau_b$, the lifetime of the bound state,
is $|2 \pi V(\epsilon_b)|^{-1}$. The oscillatory
exponential behaviour of $b(t)$ is because 
the amplitude $| \langle b | \psi_E \rangle |^2$
is a Lorentzian sharply peaked about the renormalised energy. 
The probability to remain in the bound state is therefore a pure
exponential decay, $ |b(t)|^2 = \exp( - t/\tau_b)$. This is
the usual reason that tunneling decay is exponential. 

Now consider instead a model in which two or more bound states
are coupled to the same continuum by a perturbation. Even if the
perturbation does not directly couple the two bound states, they
become effectively coupled due to their coupling to the same 
continuum. Thus if the system starts in one bound
state it will undergo damped oscillations into the other bound state(s),
the damping being produced by the tunneling decay into the continuum.

To put this idea on a quantitative footing consider, again following
Fano \cite{fano}, a model with $n$ bound states $|i\rangle$ where $i = 1, 2, 3,
\ldots n$ and a single continuum $|E\rangle$. These states are assumed
to be orthonormal and eigenstates of the unperturbed Hamiltonian $H_0$
with eigenvalues $\varepsilon_i$ and $E$ respectively. We assume that 
the only non-vanishing matrix elements of the perturbation are 
$\langle E | V | i \rangle = V_i(E)$. Once the perturbation is turned on
the bound states will dissolve into the continuum; the perturbed 
continuum states will be denoted $|\psi_E\rangle$. Once again these
states are most easily derived using the Lipmann-Schwinger method.
A key quantity in this analysis is the hermitean part of the bound state
self-energy matrix \footnote{The self-energy 
$ \Sigma_{ij} (E) \equiv  \langle i | V (E - H)^{-1} V |j \rangle$.}
\begin{equation}
F_{ij} (E) = P \int d E' \frac{V_i(E') V^{\ast}_j(E')}{E - E'}.
\label{eq:selfmatrix}
\end{equation}
Formally the oscillations in the tunneling may be traced to the
fact that the self-energy term $F_{ij}$ is not diagonal. It is useful
to determine the eigenvectors of the $n \times n$ matrix
$\varepsilon_i \delta_{ij} + F_{ij}(E)$ denoted as $A_{i \nu}(E)$ with
eigenvalues $\varepsilon_{\nu}$. If we define $V_{\nu} (E) = 
\sum_i V_i (E) A_{i \nu} (E) $ then we find that the overlap of
the perturbed continuum eigenstates $|\psi_E\rangle$ 
with the unperturbed bound states is
\begin{equation}
\sum_{i=1}^{n}
A_{i \mu}^{\ast} (E) \langle i | \psi_E \rangle = 
\frac{1}{\pi V_{\mu} (E)} \tan [ \Delta_{\mu} (E) ] \cos [ \Delta(E) ].
\label{eq:multiboundoverlap}
\end{equation}
Here the phase shifts $\Delta_{\mu}$ and $\Delta$ are defined via
\begin{equation}
\tan \Delta_{\mu} (E) = \frac{ \pi | V_{\mu}(E) |^2 }{ E - \varepsilon_{\mu}(E) };
\hspace{3mm}
\tan \Delta = \sum_{\mu=1}^{n} \tan \Delta_{\mu}.
\label{eq:multiphase}
\end{equation}
Once again the renormalised energies $\epsilon_{\mu}$ may be defined
as the values of $E$ at which $E - \varepsilon_{\mu}(E)$ vanishes. The phase
$\Delta_{\mu}$ then jumps from $\pi$ to zero as the energy is swept 
through $\epsilon_{\mu}$ from below.

Fano studied the lineshape of radiative transitions into a continuum
coupled to multiple bound states using the model and solution 
outlined above. Here we investigate what happens if the system
starts in one of the bound states $|i\rangle$. Making the same approximations
that led to eq (\ref{eq:decay}) a straightforward calculation reveals 
that the amplitude to remain in the state $|i\rangle$ is 
\begin{equation}
b_i(t) = \sum_{\mu} | A_{i \mu} (\epsilon_{\mu}) |^2 \exp( - i \epsilon_{\mu} t )
\exp \left( - \frac{t}{2 \tau_\mu} \right).
\label{eq:oscillatorydecay}
\end{equation}
Here $\tau_{\mu} = 2 \pi | V_{\mu} (\epsilon_{\mu})|$. Since the different
terms in the superposition oscillate at different frequencies they will
interfere leading to oscillations in the probability $|b_{i} (t) |^2$. 
For example if there are just two coupled bound states 
\begin{eqnarray}
|b_1(t) |^2 & = & 
| A_{11} (\epsilon_1) |^4 \exp \left( - \frac{t}{\tau_1} \right) +
| A_{12} (\epsilon_2) |^4 \exp \left( - \frac{t}{\tau_2} \right)
\nonumber \\
&  + &
2 | A_{11} |^2 |A_{12}|^2 \cos[ (\epsilon_n - \epsilon_m) ] 
\exp \left( - \frac{t}{2} [ \frac{1}{\tau_1} + \frac{1}{\tau_2} ] \right).
\nonumber \\
\label{eq:oscillations}
\end{eqnarray}
If the decay times are comparable and the oscillation frequency 
sufficiently high a departure from simple exponential decay 
should be easily observable. 

Evidently this analysis is relevant to electrons on helium in a tilted
magnetic field where there
are in fact multiple bound states that couple to the same continuum.
There are already experimental studies on the tunneling rates of
electrons on helium in a magnetic field \cite{andrei} but not from the type of resonant
states considered here. It would be desirable in future work to identify
an optimum set of coupled bound states for an experimental detection
of oscillatory exponential decay. In addition 
analogues throughout atomic, solid state, nuclear and particle
physics bear investigation. Another important extension of this
work is to consider the damping of Fano resonant effects by a bath
of oscillators. Such damping is interesting both as a matter of principle
and in order to incorporate the effect of ripplons and vapour scattering
on the single particle effects analysed here.

In summary we have shown that electrons on the liquid helium in a magnetic
field have bound states that 
dissolve into the continuum when an inplane magnetic field is 
applied; however they leave behind an imprint in the form of 
a phase shift and a sharp microwave absorption feature whose lineshape 
is tunable by varying the polarisation and the inplane magnetic field. 
In addition we point out that the tunneling decay of electrons from these 
formerly bound states may have an oscillatory exponential
form.

We acknowledge discussions with Arnold Dahm and Francesc Ferrer.

\end{document}